\documentclass[twocolumn,floats,floatfix,showpacs,amssymb,prd,superscriptaddress,nofootinbib]{revtex4-1}
\usepackage{graphicx, epsfig, amssymb} 
\usepackage{amsmath, amsfonts}
\usepackage{bm} 
\usepackage{hyperref}
\usepackage[usenames]{color}

\def\be{\begin{equation}}
\def\ee{\end{equation}}
\def\beq{\begin{eqnarray}}
\def\eeq{\end{eqnarray}}

\def\nn{\nonumber}

\newcommand{\bea}{\begin{eqnarray}}
\newcommand{\eea}{\end{eqnarray}}
\newcommand{\ben}{\begin{enumerate}}
\newcommand{\een}{\end{enumerate}}
\newcommand{\bi}{\begin{itemize}}
\newcommand{\ei}{\end{itemize}}

\begin{document}

\title{\large Anti de Sitter black holes and branes in dynamical Chern-Simons gravity:\\
perturbations, stability and the hydrodynamic modes}

\author{T\'erence Delsate} \email{terence.delsate@ist.utl.pt}
\affiliation{CENTRA, Departamento de F\'{\i}sica, 
Instituto Superior T\'ecnico, Universidade T\'ecnica de Lisboa - UTL,
Av.~Rovisco Pais 1, 1049 Lisboa, Portugal.}

\author{Vitor Cardoso} \email{vitor.cardoso@ist.utl.pt}
\affiliation{CENTRA, Departamento de F\'{\i}sica, 
Instituto Superior T\'ecnico, Universidade T\'ecnica de Lisboa - UTL,
Av.~Rovisco Pais 1, 1049 Lisboa, Portugal.}
\affiliation{Department of Physics and Astronomy, The University of Mississippi, University, MS 38677, USA.}

\author{Paolo Pani} \email{paolo.pani@ist.utl.pt}
\affiliation{CENTRA, Departamento de F\'{\i}sica, 
Instituto Superior T\'ecnico, Universidade T\'ecnica de Lisboa - UTL,
Av.~Rovisco Pais 1, 1049 Lisboa, Portugal.}

\date{\today} 

\begin{abstract}
Dynamical Chern-Simons (DCS) theory is an extension of General Relativity in which the gravitational field is coupled to a scalar field
through a parity violating term. We study perturbations of anti-de Sitter black holes and branes in such a theory, and show that the relevant equations reduce to a set of coupled ODEs which can be solved efficiently through a series expansion.
We prove numerically that black holes and branes in DCS gravity are stable against gravitational and scalar perturbations in the entire parameter space. Furthermore, by applying the AdS/CFT duality, we relate black hole perturbations to hydrodynamic quantities in the dual field theory, which is a (2+1)-dimensional isotropic fluid with broken spatial parity. The Chern-Simons term does not affect the entropy to viscosity ratio and the relaxation time, but instead quantities that enter the shear mode at order $q^4$ in the small momentum limit, for example the Hall viscosity and other quantities related to second and third order hydrodynamics. We provide explicit corrections to the gravitational hydrodynamic mode to first relevant order in the couplings.
\end{abstract}

\maketitle

\section{Introduction}
General Relativity (GR) stands as one of the most elegant physical theories, and it describes gravitational interactions from small to at least solar system scales with high accuracy. Nevertheless, there are several indications that Einstein's theory should be modified, this ``evidence'' ranging from unification to quantization arguments. A variety of different proposals to extend GR have been made ever since its conceptual formulation almost a century ago. A promising extension of GR is Dynamical Chern-Simons (DCS) gravity \cite{Deser:1982vy,Lue:1998mq,Jackiw:2003pm,Alexander:2009tp}, in which the Einstein-Hilbert action is modified by adding a parity-violating Chern-Simons (CS) term, which couples to gravity via a scalar field. This correction arises in many contexts. Such a term could help explaining several problems of cosmology, from inflation (as discussed by Weinberg \cite{Weinberg:2008mc}) to baryon asymmetry \cite{GarciaBellido:2003wd,Alexander:2004xd,Alexander:2004us}. In most of the moduli space of string theory, a CS correction is required to preserve unitarity; furthermore, duality symmetries induce a CS term in all string theories with a Ramond-Ramond scalar \cite{Polchinski:1998rr}. In loop quantum gravity, the CS term is required to ensure gauge invariance of the Ashtekar variables \cite{Ashtekar:1988sw} and it also arises naturally if the Barbero-Immirzi parameter is promoted to a field \cite{Taveras:2008yf,Mercuri:2009zt}.

DCS gravity is described by standard GR supplemented with a parity violating term, which can be thought of as the 
parity violating counterpart to the Gauss-Bonnet term: in analogy to the Gauss Bonnet term, in four dimensions the CS term is a topological term, which enters the equations of motion only when coupled to a scalar field. Due to the parity violating nature of the CS term, spherically- and plane-symmetric solutions in GR, such as the Schwarzschild or the Friedman-Robertson-Walker-Lemaitre spacetime, are still solutions of the DCS theory. Nevertheless, these spacetimes, which are simultaneously solutions in different gravity theories, respond differently in both theories to {\it generic external perturbations}: the perturbed spacetime needs not, and in general does not, obey the underlying symmetry of the background solution and therefore probes more completely the structure of the field equations. This point, which was raised by many authors (see e.g. Ref.~\cite{Barausse:2008xv}), can be clearly seen by considering perturbations of a Schwarzschild black hole in DCS theory: in GR, the linearized equations can be reduced to two decoupled second-order ODEs and both gravitational polarizations have the same characteristic spectra \cite{Berti:2009kk}; in DCS gravity however, the axial sector of the theory couples to the scalar field, resulting in two coupled ODEs for this sector, and a loss of isospectrality in the oscillations, or quasinormal modes (QNMs) \cite{Cardoso:2009pk,Molina:2010fb}. In flat spacetime, non-rotating black holes in both theories are stable~\cite{Molina:2010fb}.

For asymptotically anti-de Sitter (AdS) spacetimes, most of the previous considerations still hold. However, so far investigations of DCS gravity have been almost only limited to the case of asymptotically flat spacetime, with the exception of Ref.~\cite{Ahmedov:2010fz}, where some solutions to the non-dynamical version of CS gravity with a negative cosmological constant have been studied.  Very recently, DCS gravity in asymptotically AdS spacetime has been considered in the context of gauge/gravity correspondence~\cite{Saremi:2011ab}. 
However no study has yet been made of either the stability properties or the oscillation modes of AdS black holes/branes in DCS gravity.

AdS black hole solutions are of particular interest in relation to the celebrated AdS/CFT correspondence~\cite{Maldacena:1997re}.
This duality has been originally established between $d=4$, ${\cal N}=4$ Yang-Mills theory and a string theory in $AdS_5\times S^5$.
However, it is often seen as a particular realization of a more general \emph{gauge/gravity} duality, mapping AdS solutions of the gravity theory in $d+1$ dimensions to conformal field theories living on the $d$ dimensional boundary spacetime and viceversa. In this context, AdS black holes are dual to thermal quantum field theories at strong coupling and their characteristic oscillations dominates the near-equilibrium response of the dual gauge theory. More precisely, black hole QNMs in the bulk are related to the poles of the retarded correlators of the corresponding dual operator, providing useful insights on the transport coefficients and on the quasiparticle spectrum (see Refs.~\cite{Berti:2009kk,Kovtun:2005ev} for a review). Within Einstein's gravity in five dimensions, black hole perturbations have been extensively investigated and have revealed unexpected properties of the corresponding strongly coupled field theory. For example, the celebrated universality of the shear viscosity to entropy ratio~\cite{Kovtun:2005ev,Kovtun:2004de} can be understood in terms of the universality of black brane QNMs in the hydrodynamic limit. 

Considering the holographic duals to alternative theories of gravity is interesting for several reasons.
First of all, any consistent correction (as the DCS term) to GR can in principle modify the dynamics of the dual theory, with potentially interesting effects. There has been an intense effort to apply this approach in order to describe holographic duals of strongly coupled system from a phenomenological point of view.
Furthermore, adding second-order in curvature corrections to the gravity theory corresponds, from the holographic point of view, to consider the next to leading order corrections in a strong coupling expansion. 

Since Schwarzschild black holes are still solutions of DCS gravity, the holographic theory will be the same as in GR, but the CS coupling will affect the linear response of the dual theory at equilibrium. For example hydrodynamic quantities and quasi-particle spectra will carry the imprint of the CS parity violation. Very recently~\cite{Saremi:2011ab}, DCS gravity has been considered as a simple holographic realization of a $(2+1)$-dimensional isotropic fluid with broken spatial parity in a bottom-up approach, and it was shown that the holographic theory may possess a non-zero Hall viscosity. The Hall viscosity is related to the presence of a non-trivial background scalar field, encoding the nature of the parity violation in CS gravity.

In this work we shall follow the same bottom-up approach and investigate the perturbations spectrum of Schwarzschild-anti-de Sitter (SAdS) black holes and branes in four dimensions. The $(2+1)$-dimensional dual field theory would be defined on a two dimensional sphere (or plane, for black branes) and it will be free of triangle anomalies~\cite{Bhattacharyya:2008jc,Erdmenger:2008rm}. Because the SAdS background is even-parity, the Hall viscosity studied in Ref.~\cite{Saremi:2011ab} vanishes, at least at first order. Nevertheless, the CS coupling will affect the hydrodynamic limit of the dual theory, in a characteristic way which we explicitly compute.
Our main results are that 

\noindent {\bf (i)} black holes and branes in this theory are stable.

\noindent {\bf (ii)} in the hydrodynamical limit, the shear mode acquires DCS corrections. We find
\be
\omega_\text{shear} \sim - i\frac{q^2}{3r_h} - i\left( \gamma\frac{L^2}{r_h^3} - \chi\frac{\alpha^2}{L^2r_h^3} \right)q^4 + {\cal O}(q^6)\,.
\ee
where $\gamma=\frac{9-9\log 3+\sqrt{3}\pi}{162}\sim0.0281$ and the novel correction factor $\chi$ is
\be
\chi=\frac{3}{640}\left(201-20\sqrt{3}\pi-60\log3\right)\sim0.123072\,.
\ee
Here, $\alpha$ is a DCS coupling (defined below), $r_h,L$ are the horizon and the AdS radius respectively.

\noindent {\bf (iii)} the sound mode is unaffected. In particular, the ratio $\eta/s=1/4\pi$ still holds in DCS gravity.
\section{Setup}
DCS gravity with a negative cosmological constant is described by the following action
%
\bea
S&&=\kappa\int d^4x\sqrt{-g}(R - 2\Lambda)- \frac{\beta}{2}\int d^4x \sqrt{-g}\nabla_a\phi\nabla^a\phi +\nn\\
&& + \frac{\alpha}{4}\int d^4x\sqrt{-g}\phi\,{}^*R R\,,
\label{action}
\eea
%
where $\phi$ is a real scalar field, which couples to the Pontryagin density 
\be
^*R R = \frac{1}{2}R_{abcd}\epsilon^{baef}R^{cd}_{\;\;\;ef}\,.
\ee
Without loss of generality, we set $\kappa=1=\beta$ and write the cosmological constant as $\Lambda = -3/L^2$, where $L$ is the AdS curvature radius. The variation of the action \eqref{action} with respect to the scalar field and to the metric component leads to the following system of coupled equations:
\bea
R_{ab}&=& -\alpha C_{ab} + \frac{1}{2} \nabla_a\phi\nabla_b\phi,\\
\Box\phi &=&  -\frac{\alpha}{4}\,{}^*RR,
\eea
where 
\be
C^{ab}=\nabla_c\phi\epsilon^{cde(a}\nabla_eR^{b)}_{~~d}
+\nabla_c\nabla_d\phi\,^*R^{d(ab)c}\,.\label{Ctensor}
\ee

We consider spherically- and plane-symmetric black objects, which we write in a unified form as \cite{Lemos:1994xp}
\be
ds^2 = g_{ab}dx^adx^b = -f(r)dt^2 + \frac{dr^2}{f(r)} + r^2 d\Sigma_\kappa^2, 
\label{metric}
\ee
where $f(r) = r^2/L^2 + \kappa - 2M/r$, and $d\Sigma_1^2 = d\theta^2 +\sin^2(\theta)d\varphi^2$ with $\theta\in[0,\pi],\ \varphi\in[0,2\pi]$ for the spherically symmetric ($\kappa=1$), whereas $d\Sigma_0^2 = d\theta^2 + d\varphi^2$ with $\theta,\varphi\in ]-\infty,\infty[$ for the planar case ($\kappa=0$) respectively. The plane symmetric black hole can also be seen as a particular case of the spherically symmetric one, in the limit that the horizon is much larger than the AdS radius.

As previously discussed, spherically or plane-symmetric solutions of GR are solutions of DCS gravity. 
More precisely, for these highly symmetric spacetimes, $^*R R\equiv0\equiv C_{ab}$.
It follows that the SAdS spacetime is a solution to DCS gravity. This is also the case for black
 branes (sometimes called planar black holes)~\cite{Lemos:1994xp,Ahmedov:2010fz}. In fact, the theory reduces to Einstein-Klein 
Gordon theory with a (negative) cosmological constant and it is well known that such a theory does not admit scalar hairy black holes 
\cite{Torii:2001pg}. As a consequence, the scalar field reduces to zero and we are left with pure GR with a cosmological constant.
The situation can be different if a scalar self potential $V(\phi)$ is introduced in the action~\cite{Saremi:2011ab}. 
However in our case, even if the background solutions are the same as in GR, the dynamics of perturbations of such spacetimes can be modified due to the CS term. 

\section{Gravitational perturbations of SAdS black holes/branes in DCS gravity}
\subsection{The equations}
Gravitational perturbations of Schwarzschild black holes were considered in Refs.~\cite{Cardoso:2009pk,Molina:2010fb} in the context of DCS gravity in asymptotically flat spacetime. There, the authors showed that the metric perturbations fall in two categories: (i) a polar (even-parity) or Zerilli sector, which is unaffected by the CS coupling, and whose equations of motion are identical to GR; and (ii) the axial (odd-parity) or Regge-Wheeler, which is modified in DCS gravity by the coupling to the dynamical scalar field. The procedure carries over to the case of asymptotically AdS spacetimes. In the following we focus only on the axial sector, since the polar sector is identical to the GR case (see e.g. Ref.~\cite{Berti:2009kk}).
We parametrize the metric perturbations around the SAdS background given by \eqref{metric} adopting the Regge-Wheeler gauge as in Ref.~\cite{Cardoso:2009pk},
\be
ds^2 = g_{ab}dx^adx^b + 2 h_0(r)\mathbb T dt + 2 h_1(r)\mathbb T dr\,,
\ee
where
\be
\mathbb T = -\frac{1}{F_\kappa(\theta)}\partial_\varphi Y_{lm}^\kappa d\theta + F_\kappa(\theta) \partial_\theta Y_{lm}^\kappa d\varphi\,,
\ee
and $F_\kappa(\theta) = \sin(\theta)$ for $\kappa=1$, $F_\kappa(\theta)=1$ for $\kappa=0$, $Y_{lm}^{\kappa}$ are the scalar harmonics on the sphere (plane) for $\kappa=1$ ($\kappa=0$). Due to the background symmetry, we can fix $m=0$ without loss of generality.

The perturbed gravitational equations then reduce to a set of three coupled equations:
\bea
E_1&=&\frac{2 r h_1}{L^2}+\frac{i \omega h_0}{f}+f(r) h_1'+\frac{2 M h_1}{r^2}=0\,,\nonumber\\
E_2&=&\frac{1}{2} f h_0''-\frac{h_0 (2 f+(q-\kappa ) (q+2 \kappa ))}{2 r^2}+\frac{1}{2} i \omega  f h_1'\nonumber\\
&+&\frac{i \omega f h_1}{r}+\frac{6 M \alpha f\sigma}{r^5}-\frac{3 M \alpha  f \sigma'}{r^4}=0\,,\nonumber\\
E_3&=&-\frac{i \omega  h_0'}{2 f}+\frac{i \omega h_0}{r f}+\frac{h_1\left(r^3 \omega ^2-r f (q-\kappa ) (q+2\kappa )\right)}{2 r^3   f}\nonumber\\
&&+\frac{3 i M \alpha  \omega  \sigma}{r^4 f}=0\,,
\eea
where $\sigma=\phi(r)r$. Here and in the following, when $\kappa=1$, $q$ is an integer ($q=l=2,3,...$), whereas, for $\kappa=0$, $q$ is real and we write $q=\sqrt{q_x^2+q_y^2}$, in terms of the components $(q_x,\,q_y)$ of the spatial momentum on the plane, the scalar harmonics on the plane simply being $\sim \exp[i(q_x \theta+q_y\varphi)]$.
The three equations above are not independent and obey the following relation 
\be
\left(-\frac{2 i r^2 f E_3}{\omega }\right)' + \frac{2 r^2 E_2}{f} + \frac{i(q-\kappa )(q+2 \kappa )E_1}{\omega } =0\,.\nn
\ee
We supplement the equations by the derivative of $E_1$ and solve the resulting set of four equations for $h_0,h_0',h_0''$ and $h_1$.
Defining the Regge-Wheeler master function $Q(r)$ as 
\be
h_1 = \frac{r Q}{\omega f}\,,
\ee
the equations for $Q$ and $\sigma$ reduce to
\bea
&&\frac{d^2}{dr_*^2}Q + (\omega^2+V_{RW}) Q +T_{RW}  \sigma=0\,,\label{coupled_eqs1}\\
&&\frac{d^2}{dr_*^2}\sigma +(\omega^2 + V_{S}) \sigma +T_{S} Q=0\,,
\label{coupled_eqs2}
\eea
where $r_*$ is the tortoise coordinate defined as $dr/dr_*=f(r)$ and
\bea
V_{RW}&=&  - f\left( \frac{q(q+\kappa)}{r^2}-\frac{6M}{r^3} \right),\ T_{RW}=\frac{6 i M f \alpha }{r^5}\,,\nonumber\\
V_{S}&=&  -f\left( \frac{q(q+\kappa)}{r^2}\left( 1+\frac{36\alpha^2 M^2}{r^6} \right)+\frac{2M}{r^3}+\frac{2}{L^2} \right)\,,\nonumber\\
T_{S}&=& (q+2\kappa)(q+\kappa)q(q-\kappa)\frac{6 i\alpha fM}{ r^5}\,.
\eea
These equations form the basis of the present analysis and they reduce to the correct asymptotically flat case \cite{Cardoso:2009pk,Molina:2010fb} when $L\to\infty$. Close to the horizon ($r_*\rightarrow-\infty$), the asymptotic solutions to this system are
\bea
Q&\to& Q_H^{out}e^{i\omega r_*}+Q_H^{in}e^{-i\omega r_*},\nonumber\\
\sigma&\to& \sigma_H^{out}e^{i\omega r_*}+\sigma_H^{in}e^{-i\omega r_*}\,.
\eea

Close to spatial infinity ($r^*\rightarrow0$) on the other hand, we have
\bea
Q&=& \frac{Q_{\infty}^{\rm reg}}{r}+Q_{\infty}^{\rm irr}\,,\nonumber\\
\sigma &=& \frac{\sigma_{\infty}^{\rm reg}}{r^2} + \sigma_{\infty}^{\rm irr} r\,.
\eea
Regular scalar perturbations should have $\sigma_{\infty}^{\rm irr}=0$, corresponding to
Dirichlet boundary conditions at infinity. The case for
gravitational perturbations is less clear and the requirement of regularity is not unique.
However, most calculations in the literature assume Dirichlet boundary
conditions also for the gravitational perturbations and this is the choice we adopt here
(see Ref.~\cite{Berti:2009kk} and references therein for a more detailed discussion on this problem).
Therefore, we impose the following boundary conditions:
\be
Q_{H}^{out}=\sigma_{H}^{out}=Q_{\infty}^{\rm irr}=\sigma_{\infty}^{\rm irr}=0\,.\label{BCs}
\ee
In relation to the gauge/gravity correspondence, it has been shown that such boundary conditions allow one to recover the shear mode, but not the sound mode \cite{Miranda:2008vb,Morgan:2009pn}. However, the sound mode is retrieved from the polar sector which is unaffected by the DCS coupling, 
and we therefore consider once and for all Dirichlet conditions at spatial infinity.
\subsection{Frobenius expansion}
In asymptotically flat spacetime, solving the coupled system analogous to Eqs.~\eqref{coupled_eqs1}-\eqref{coupled_eqs2} above is not an easy task~\cite{Molina:2010fb}. However, in an asymptotically AdS spacetime, due to the boundary conditions~\eqref{BCs}, the coupled system~\eqref{coupled_eqs1}-\eqref{coupled_eqs2} can be solved by extending the series expansion method, used in Refs.~\cite{Horowitz:1999jd,Cardoso:2001bb} to study SAdS black holes in GR.

We start by re-writing $f$ as 
\be
f(r) = \frac{(r-r_h) \left(\kappa L^2+r^2+r r_h+r_h^2\right)}{L^2 r}\,,
\ee
where $r_h+r_h^3/L^2 = 2M$. We factorize the horizon behavior according to
\bea
Q(r)=e^{-i \omega r_*}\tilde Q(r)\,,\qquad \sigma(r)=e^{-i \omega r_*}\tilde \sigma(r)\,,
\label{factorization}
\eea
and we introduce a new variable $x=1/r$, rewriting the relevant equations~\eqref{coupled_eqs1}-\eqref{coupled_eqs2} as 
\bea
&&\left[(x-x_h)s_1(x) \frac{d^2}{dx^2}  + f_1(x) \frac{d}{dx} + \frac{v_{RW}(x)}{x-x_h} \right]\tilde Q(x)+\nn\\
&& + \frac{t_{RW}(x)}{x-x_h}\tilde \sigma(x)=0,\nonumber\\
&&\left[(x-x_h)s_2(x) \frac{d^2}{dx^2}  + f_2(x) \frac{d}{dx} + \frac{v_{S}(x)}{x-x_h} \right]\tilde \sigma(x)+\nn\\
&& + \frac{t_{S}(x)}{x-x_h}\tilde Q(x)=0\,,\nonumber
\eea
where we have defined
\beq
x_h&=& \frac{1}{r_h}\,,\quad s_1(x) = s_2(x) = \frac{x^4 f}{x-x_h}\,,\\
f_1&=&f_2= 2 x^3 f - x^2f' -2 i\omega x^2\,,\\
v_{RW}&=&(x-x_h)\frac{V_{RW}}{f}\,,\quad t_{RW} = (x-x_h)\frac{T_{RW}}{f}\,,\\
v_{S} &=& (x-x_h)\frac{V_{S}}{f}\,,\quad t_{S} = (x-x_h)\frac{T_{S}}{f}\,. 
\eeq
Expanding the unknown functions as 
\be
\tilde Q(x) = \sum_{k=0}^\infty q_k\left(x-\frac{1}{r_h}\right)^k\,,\quad \tilde\sigma(x) = \sum_{k=0}^\infty \sigma_k \left(x-\frac{1}{r_h}\right)^k\,,\nn
\ee
we find the modified recursion relations
\begin{widetext}
\bea
q_n&=& -\frac{1}{P^1_n}\sum_{k=0}^{n-1}\left[\left( k(k-1)s_{1,n-k} + k f_{1,n-k} + v_{RW,n-k}\right)q_k + t_{RW,n-k}\sigma_k \right],\nonumber\\
\sigma_n&=& -\frac{1}{P^2_n}\sum_{k=0}^{n-1}\left[\left( k(k-1)s_{2,n-k} + k f_{2,n-k} + v_{S,n-k}\right)\sigma_k +t_{S,n-k}q_k \right],
\eea
\end{widetext}
where $P^i_n = n(n-1)s_{i,0} + n f_{i,0}$ and where we expanded the coefficients of the equation as
\be
X (x) = \sum_{k=0}^\infty X_k(x-x_h)^k\,,
\ee
$X$ generically denoting $s_1,s_2,f_1,f_2,v_{S},t_{S},v_{RW},t_{RW},\tilde\sigma,\tilde Q$.

There are two free parameters in the expansion: $q_0$ and $\sigma_0$, and all the coefficients may depend on $\omega$, so we have essentially $(2+1)$ parameters. Either $q_0$ or $\sigma_0$ can be arbitrarily chosen due to the linearity of the equations~\eqref{coupled_eqs1}-\eqref{coupled_eqs2}. Thus, we are left with two parameters, $\sigma_0$ (or $q_0$) and $\omega$, which can be found by simultaneously requiring $Q(x\to0) = 0$ and $\sigma(x\to0)=0$.  
The above procedure can be used to obtain the characteristic frequencies $\omega$ and the ratio $\sigma_0/q_0$.

\section{Numerical Results}
We solved the system of equations~\eqref{coupled_eqs1}-\eqref{coupled_eqs2} using the series expansion method described above.
For numerical purposes, the series has to be truncated at some finite order $N$. As a first check on the numerical results,
we verified that the corresponding solution $\omega_N$ converges for large or moderately large $N$.
A typical convergence plot is shown in Fig.~\ref{fig:conv_sph} in the appendix, for two particular sets of data and for the spherically symmetric case. Usually the spherically symmetric case  requires a larger truncation order than that required for convergence in the planar case.
This is analogous to what happens in GR: the series method works well for large and intermediate black holes~\cite{Horowitz:1999jd,Berti:2009kk}, but it ceases to converge for small black holes. In the latter case, other methods (for example the resonance method~\cite{Berti:2009wx}) can be applied, but they do not extend naturally to coupled systems of equations as the one we deal with here.

The truncation order needed to reach convergence strongly depends on the parameters. Roughly speaking, 
for a fixed accuracy the truncation order increases with $\alpha$.
For instance, our numerical experiments show that if $\kappa=0$, $N\approx 30 - 50$ to achieve an accuracy of a few tenths of a percent for any $\alpha/r_h^2\lesssim5$. On the other hand, for spherical black holes with $\kappa=1$, $N\approx 60$ is required already for $\alpha/r_h^2\sim1$, this number increases for smaller spherical black holes. In what follows, we systematically check consistency of the last points by increasing the truncation order.

Note that, even if the limit where $L\rightarrow\infty$ leads to the same system of equations than Ref.~\cite{Molina:2010fb}, the problem is not equivalent, due to the boundary conditions. As a consequence, it is not possible to recover results for the flat case. 
When $\alpha=0$, we recover the well-known modes of pure GR with a cosmological constant~\cite{Cardoso:2001vs,Berti:2009kk}. Finally, in the hydrodynamic limit at small coupling, we found an analytic expression of the hydrodynamic mode, as described below.
Our numerical results perfectly agree with the analytical result.
\subsection{QNM spectrum of AdS Schwarzschild black branes in DCS gravity}
%
\begin{figure*}[ht]
\begin{center}
\begin{tabular}{cc}
\epsfig{file=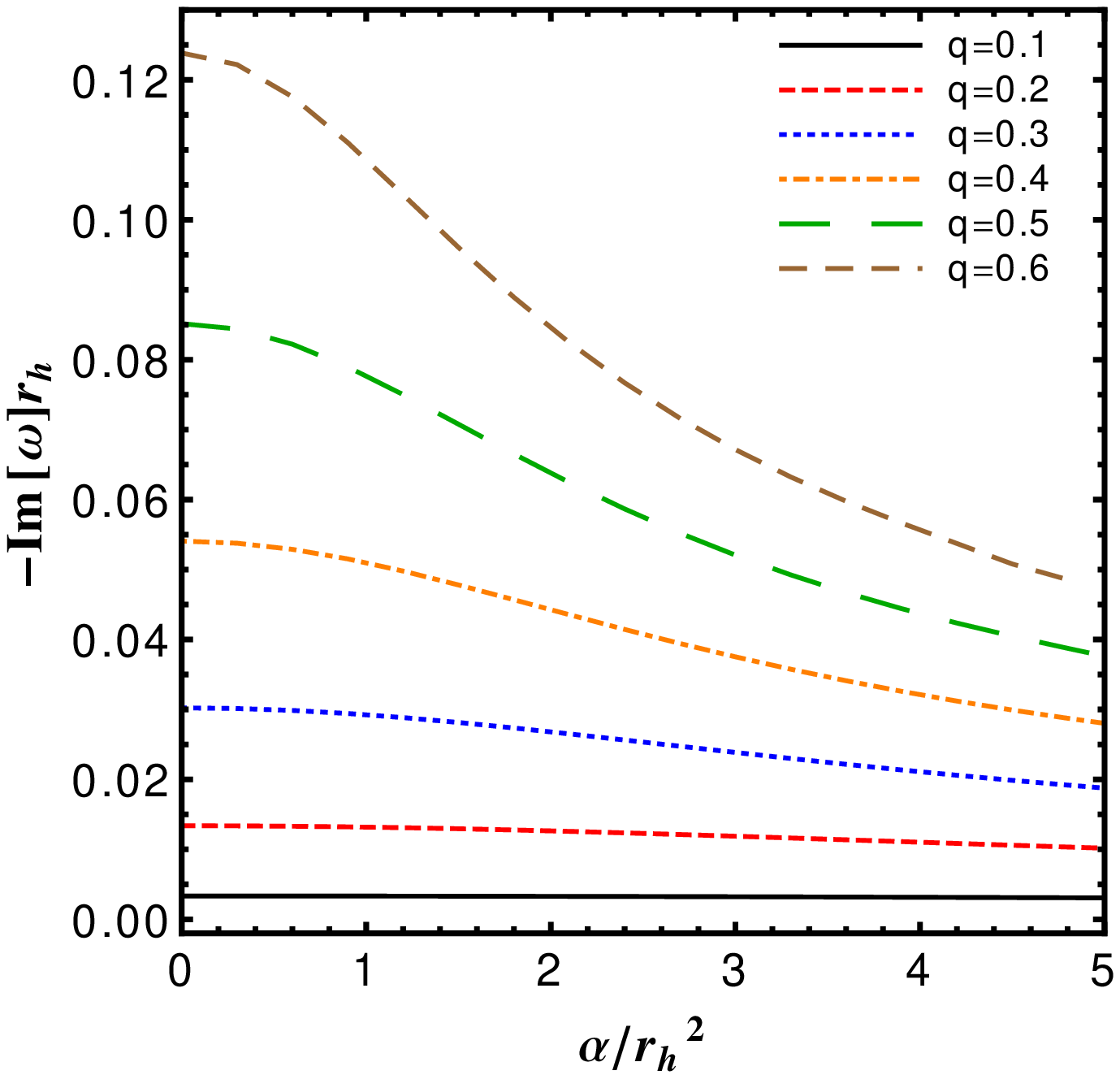,height=6.6cm,angle=0}&
\epsfig{file=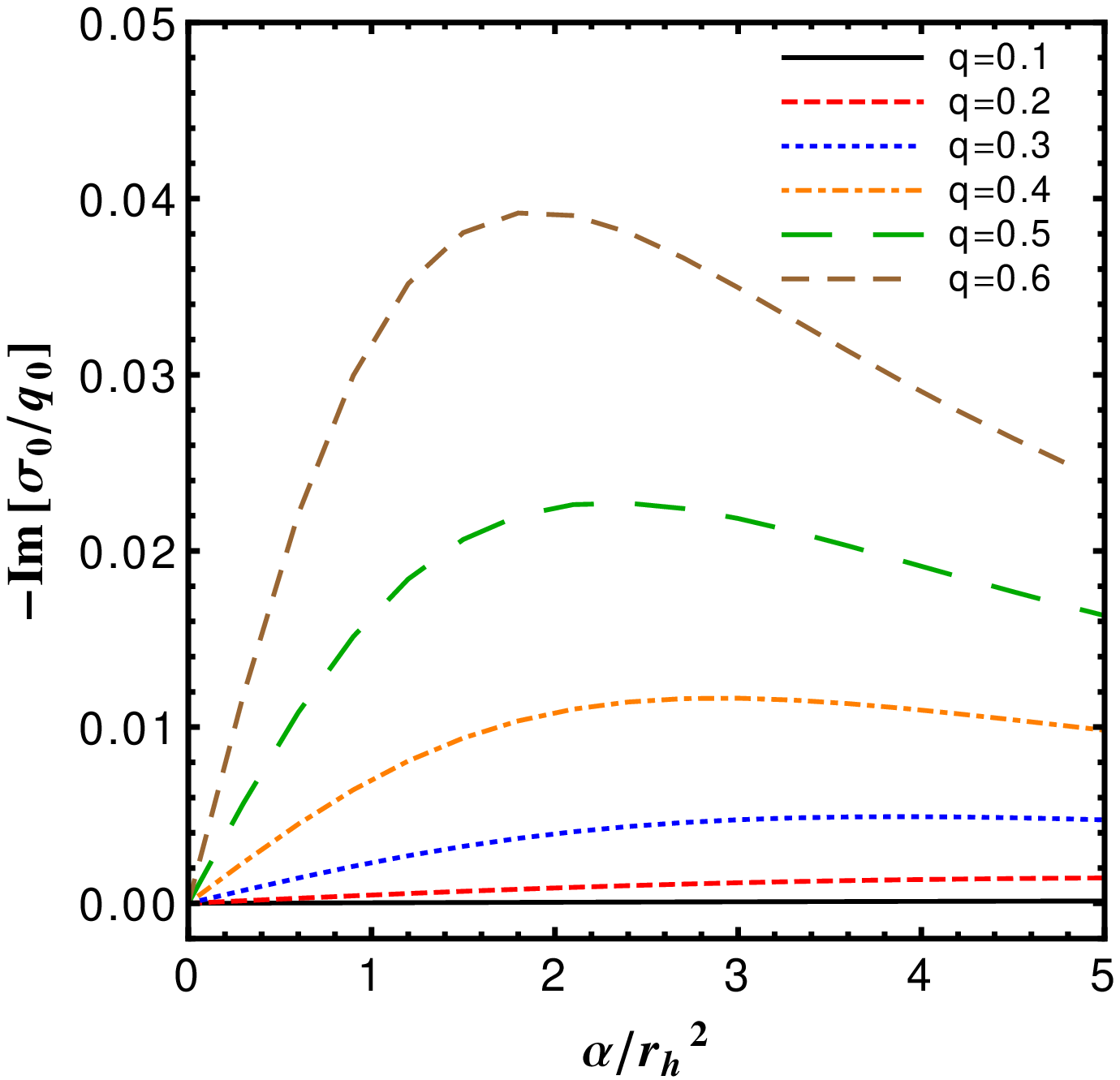,height=6.65cm,angle=0}
\end{tabular}
\caption{Hydrodynamic mode for Schwarzschild black branes as a function of the DCS coupling $\alpha$, for different values of the momentum $q$.
\label{fig:hydro2}}
\end{center}
\end{figure*}
\begin{figure*}[ht]
\begin{center}
\begin{tabular}{cc}
\epsfig{file=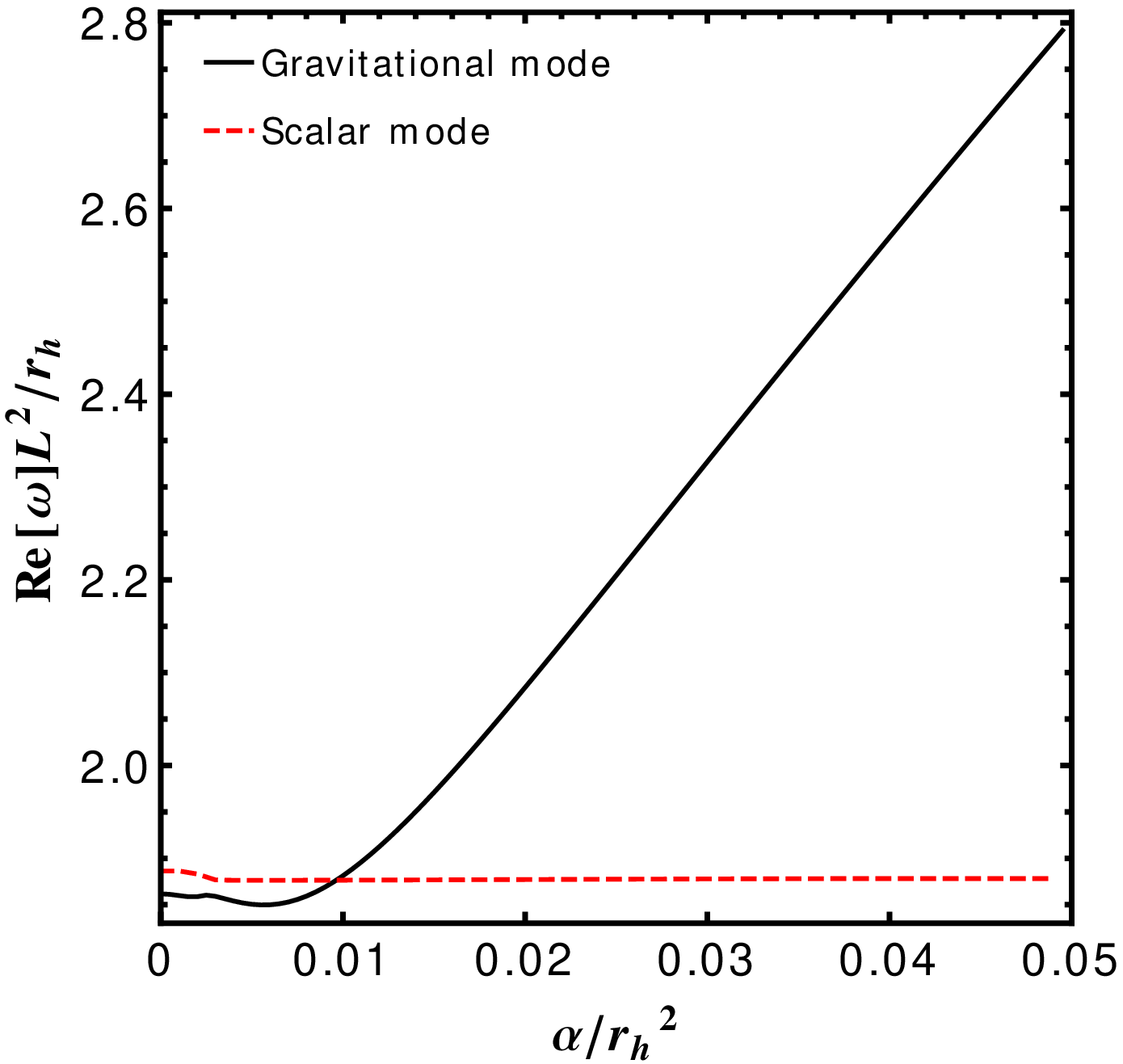,height=6.6cm,angle=0}&
\epsfig{file=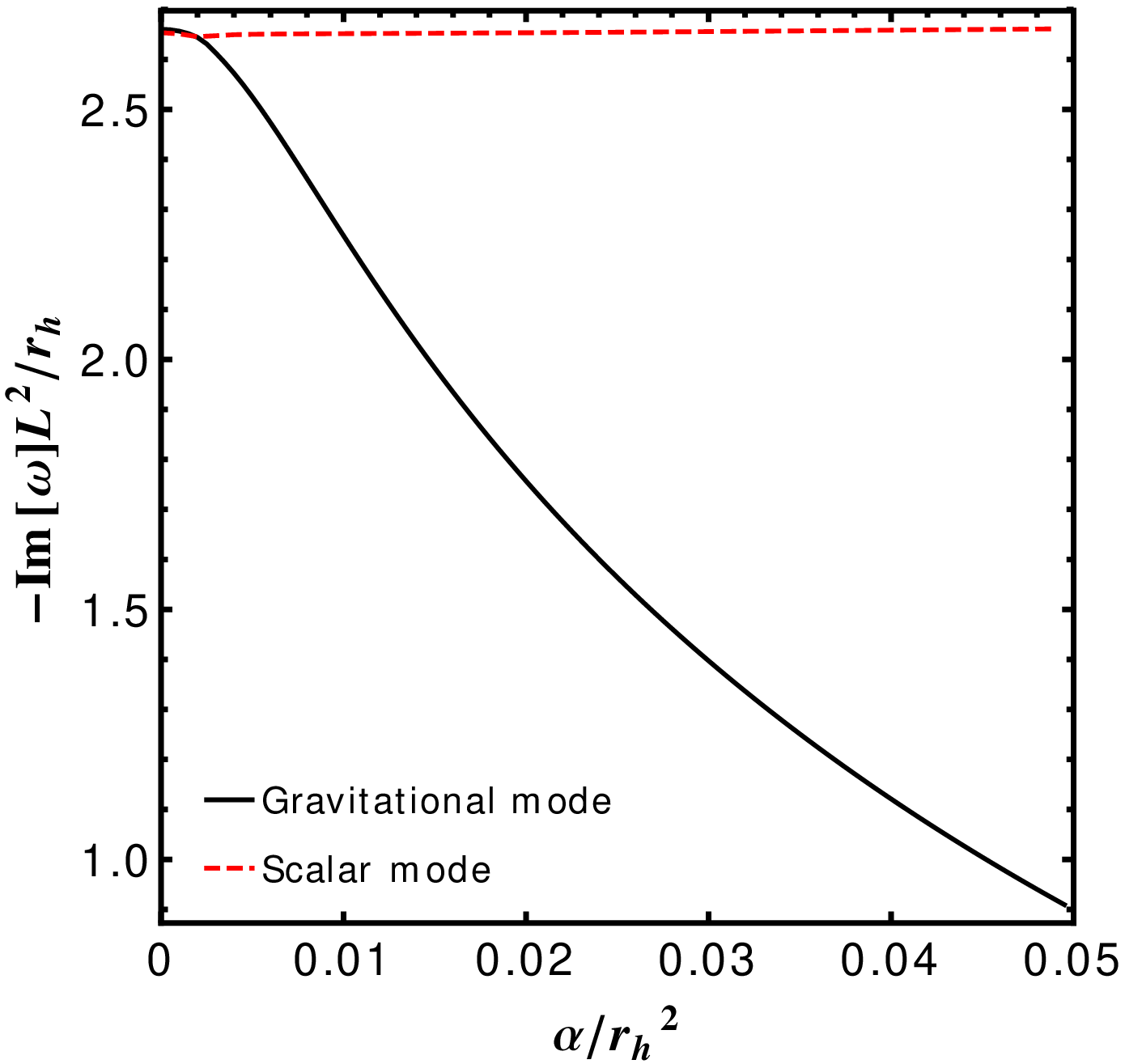,height=6.5cm,angle=0}
\end{tabular}
\caption{First overtones for scalar and gravitational modes as functions of $\alpha$ for $q=3$. The scalar mode is roughly independent from $\alpha$. Different values of $q$ give the same qualitative result.
\label{fig:grav+scal}}
\end{center}
\end{figure*}
Using the method described above, we computed the QNMs of a Schwarzschild black brane for different values of $\alpha$ and $q$. 
Our results are summarized in Figs.~\ref{fig:hydro2} and \ref{fig:grav+scal}.
For axial gravitational perturbations and scalar perturbations (these modes are coupled in DCS gravity) of Schwarzschild-AdS black holes, the fundamental modes are purely imaginary. In the left panel of Fig.~\ref{fig:hydro2} we show their dependence on $\alpha$ and on $q$. 
As expected, the CS corrections strongly depends on $q$. Indeed, modes with $q=0$ are plane-symmetric and the parity-violating CS contribution is vanishing in this case.
In the right panel of Fig.~\ref{fig:hydro2} we also show the ratio $\sigma_0/q_0$, which is purely imaginary in this case. The ratio is vanishing when $\alpha=0$, signaling that this mode belongs to the gravitational sector. In the large $\alpha$ limit the ratio asymptotically vanishes, while it is finite for intermediate values of $\alpha$. As discussed in Ref.~\cite{Molina:2010fb}, the coupled system of equations can be understood in terms of a forced oscillator: the ratio $\sigma_0/q_0$ is related to the relative amplitude between the gravitational and the scalar waveform.

In Fig.~\ref{fig:grav+scal} we show the first overtones for gravitational and scalar QNMs. Here and in the following, the labels ``scalar'' and ``gravitational'' refer to the sector that is excited in the GR ($\alpha\to0$) limit, but in DCS gravity \emph{both} modes are proper oscillations of \emph{both} gravitational and scalar perturbations.  When $\alpha=0$, these modes are very similar, but they behave quite differently as functions of $\alpha$. Indeed, the scalar mode is approximately constant, while the real part of the gravitational mode grows linearly in the large $\alpha$ limit, while (although not shown in the plot) its imaginary part asymptotically vanishes. 
In Fig.~\ref{fig:grav+scal} we show results for $q=3$, but other values of $q$ give the same qualitative results.
Notice that in the large $\alpha$ limit the ratio $\mbox{Re}[\omega]/\mbox{Im}[\omega]\gg1$ and the series method converges very slowly. Last points in Fig.~\ref{fig:grav+scal} have been computed using $N\sim 80$ as truncation order for the series. This prevents a tracking of the modes for larger values of $\alpha$. Nevertheless, for some additional points, we increased the truncation order and we explicitly checked that the gravitational mode vanishes asymptotically. In particular, in the region of the parameter space we spanned, i.e. up to $\alpha/r_h^2\lesssim1$, none of the modes cross the real axis, i.e. we did not find any unstable mode also in the large $\alpha$ and large $q$ limit.
\subsubsection{Hydrodynamic limit and AdS/CFT correspondence}
Having a large damping time, the purely imaginary mode dominates the thermalization timescale and the hydrodynamic limit of the $2+1$ dual field theory~\cite{Kovtun:2005ev}. Indeed, in the small-$q$ limit, the location of this mode can be related to some hydrodynamic quantities (see for example Ref.~\cite{Baier:2007ix}). 

For any $d-$dimensional conformal field theory, the dispersion relations for the shear mode and for the sound mode in the hydrodynamic limit read~\cite{Natsuume:2008ha}
\bea
\omega_\text{shear}&=&-i D_\eta q^2-i D_\eta^2(\tau_\pi+\tau_3)q^4+{\cal O}(q^6) \label{omegaSHEARqft}\,,\\
\omega_\text{sound}&=&v_s q-i\frac{d-2}{d-1}D_\eta q^2+\nn\\
&+&\frac{d-2}{2(d-1) v_s}D_\eta\left(2v_s^2\tau_\pi-\frac{d-2}{d-1}D_\eta\right)q^3+{\cal O}(q^4)\,,\nn\\ \label{omegaSOUNDqft} 
\eea
where $D_\eta=\eta/(\epsilon+p)$ and $\eta$, $\epsilon$, $p$ and $v_s$ are the shear viscosity, energy density, pressure and speed of sound, respectively, as computed from first order hydrodynamics. The dispersion relations above also get contributions from higher order hydrodynamics: the relaxation time $\tau_\pi$ is a second order quantity, whereas $\tau_3$ schematically denotes all possible other contributions at order $q^4$. In GR, these contributions come only from third-order hydrodynamics, which affect the shear mode at $q^4$ order~\cite{Baier:2007ix}. However, in DCS gravity also first order hydrodynamic quantities, like the Hall viscosity~\cite{Saremi:2011ab}, may affect the shear mode at the same order.

On the other hand, the AdS/CFT duality relates the dispersion relations Eqs.~\eqref{omegaSHEARqft} and~\eqref{omegaSOUNDqft} to the QNMs of SAdS black branes in the hydrodynamic limit. These frequencies can be computed analytically in the small $q$ and small $\omega$ limit. For example for four dimensional black branes we have~\cite{Berti:2009kk,Miranda:2008vb,Morgan:2009pn}
\bea
\omega_\text{shear}&\sim& - i\frac{q^2}{3 r_h} - i\,\gamma\frac{L^2}{r_h^3}q^4 + {\cal O}(q^6) \label{omegaSHEARgravity}\,,\\
\omega_\text{sound}&\sim& \pm\frac{q}{\sqrt{2}L}-i\frac{1}{6r_h}q^2+\nn\\
&+&\frac{\sqrt{2}L}{9 r_h^2}\left[\frac{7}{9}-\frac{\log3}{2}+\frac{\sqrt{3}\pi}{2}\right] q^3 +{\cal O}(q^4) \label{omegaSOUNDgravity}\,,\qquad 
\eea
where $\gamma=\frac{9-9\log 3+\sqrt{3}\pi}{162}\sim0.0281$.
Comparing Eqs.~\eqref{omegaSHEARgravity} and~\eqref{omegaSOUNDgravity} with Eqs.~\eqref{omegaSHEARqft} and~\eqref{omegaSOUNDqft} for $d=3$, we can extract the hydrodynamic quantities. More precisely, the coefficients of $q^2$ in Eqs.~\eqref{omegaSHEARgravity} and~\eqref{omegaSOUNDgravity} are related to the universal value of $\eta/s$~\cite{Kovtun:2004de,Berti:2009kk}. The coefficient of $q^4$ in the sound mode~\eqref{omegaSOUNDgravity} can be used to extract the relaxation time $\tau_\pi$ in Eq.~\eqref{omegaSOUNDqft}. Finally, the coefficient of $q^4$ in the sound mode~\eqref{omegaSHEARgravity} can be used to extract $\tau_3$.

For small CS coupling, we can extend the approximate analytical methods used in Refs.~\cite{Natsuume:2007ty,Natsuume:2008ha,Policastro:2002se} to our coupled system. First, we factorize the wavefunctions as in 
Eqs.~\eqref{factorization}. Then, anticipating the dispersion relation for the shear mode, we write
\be
\omega=\omega_2 q^2+\omega_4 q^4+\omega_\alpha q^4\alpha^2+...\,,
\ee
which holds in the hydrodynamic limit, $q\ll1$, and for small coupling. Finally, we solve the coupled system~\eqref{coupled_eqs1}-\eqref{coupled_eqs2} order by order imposing a series expansion of the following form:
\bea
\tilde Q(r)&=&\tilde Q_0(r)+\tilde Q_2(r)q^2+\tilde Q_4(r)q^4+\tilde Q_\alpha(r)q^4\alpha^2\,,\\
\tilde \sigma(r)&=&\tilde \sigma_\alpha(r)q^4\alpha\,.\label{sigmatilde}
\eea
It turns out that the ansatz above is consistent with the equations of motion and other possible terms (like those proportional to $\sim\alpha q^2$) vanish. The constants $\omega_i$ are found order by order by imposing Dirichlet boundary conditions on the solutions $\tilde Q_i(r)$ and $\tilde \sigma_i(r)$~\cite{Natsuume:2007ty}. 
Remarkably, in this limit the coupled equations can be solved analytically. At order $\alpha q^4$, the scalar equation for $\tilde\sigma_\alpha$ is sourced by $\tilde Q_0\propto 1/r$. This can be solved analytically and its solution enters in the gravitational equation for $\tilde Q_\alpha$ as a source term. The forms of $\tilde Q_i(r)$ and $\tilde \sigma_i(r)$ are cumbersome and not particularly interesting. Nevertheless, the analytical expressions for $\omega_i$ can be obtained in simple form. The first terms, $\omega_2$ and $\omega_4$, are not affected by $\alpha$ and they are simply given in Eq.~\eqref{omegaSHEARgravity}, while the novel term $\omega_\alpha$ reads
\be
\omega_\alpha\equiv\chi\,i=\frac{3}{640}\left(201-20\sqrt{3}\pi-60\log3\right)\,i\sim0.123072i\,,
\ee
which is perfectly consistent with a fit of our numerical results shown in Fig.~\ref{fig:hydro2}.
Therefore, the shear mode of a Schwarzschild black brane in DCS gravity, in the hydrodynamic limit and for small CS coupling, reads 
\be
\omega_\text{shear} \sim - i\frac{q^2}{3r_h} - i\left( \gamma\frac{L^2}{r_h^3} - \chi\frac{\alpha^2}{L^2r_h^3} \right)q^4 + {\cal O}(q^6)\,.\label{shearfit}
\ee
On the other hand, the sound mode belongs to the polar sector and it is not affected by the CS coupling.
Our results imply that $\eta/s=(4\pi)^{-1}$ also in DCS gravity. Moreover, since the sound sector is the same as in GR, also second-order hydrodynamic quantities like the relaxation time $\tau_\pi$ (see e.g. Ref.~~\cite{Moore:2010bu}) are unaffected. The first non-vanishing correction appears in the $q^4$ coefficient of the shear mode, which includes the Hall viscosity and other quantities related to second and third order hydrodynamics. 
Comparing Eq.~\eqref{shearfit} with the dispersion relations~Eqs.~\eqref{omegaSHEARqft} and~\eqref{omegaSOUNDqft} , we obtain
%
%
%
\be
\tau_3=-\tau_\pi+\frac{27\gamma}{4\pi T}\left[1-\frac{\chi}{\gamma\,L^4}\alpha^2\right]\,,\label{tau3}
\ee
where we have set $r_h=4\pi T/3$, $\epsilon+p=sT$ and the relaxation time,
\be
\tau_\pi=\frac{18-9\log 3+\sqrt{3}\pi}{24\pi T}\,,
\ee
has the same value as in pure GR~\cite{Natsuume:2008ha}. 

Notice that, although in our case the background scalar field is vanishing, in our perturbed solution~\eqref{sigmatilde} the scalar field $\sim\alpha q^4$. This implies that the first non-vanishing correction to the Hall viscosity, $\eta_A\sim\alpha^2 q^4$ (cf. Eq.~(40) in Ref.~\cite{Saremi:2011ab}), i.e. at the same order as $\tau_3$.

We remark that the universality of $\eta/s$ has been proved for theories without higher order derivatives. Indeed, in gravity theories with fourth or higher order derivatives $\eta/s$ is not universal~\cite{Banerjee:2009fm,Buchel:2008vz,Banerjee:2009wg} and the dual field theory has a different hydrodynamics, even at first order. Moreover, corrections to first order hydrodynamics can also come from Gauss-Bonnet terms in the gravitational action~\cite{Brigante:2007nu}. Thus, it is remarkable that the CS coupling, although introducing higher curvature corrections, leaves $\eta/s$ and $\tau_\pi$ unchanged, while affecting only quantities related to the parity violation, entering the shear mode at order $q^4$ in the hydrodynamic limit.

%
%
\section{Conclusion}
In this paper, we studied gravitational and scalar perturbations of planar and spherically symmetric Schwarzschild-AdS black holes embedded in dynamical Chern-Simons gravity. By generalizing a well-known series expansion technique to compute characteristic frequencies in asymptotic AdS spacetimes to systems of coupled equations, we were able to study stability and oscillation modes of black hole and branes in this theory.
We found that these solutions are stable also in DCS gravity, even in the large coupling and large momentum limit.
In the opposite, hydrodynamic regime (small coupling and small momemntum), we explicitly computed the modification arising from the CS term to first relevant order in the couplings. By applying the gauge/gravity correspondence, we found that the entropy to viscosity ratio has the same ``universal'' value as in GR. Also the relaxation time is unaltered, while the CS coupling affects the shear mode at order $q^4$, i.e. it affects the Hall viscosity and other second order and third order quantities related to parity violation. This constitutes one of the main results of this paper. It would be highly desirable to extend the analysis done in Ref.~\cite{Saremi:2011ab} in order to compute higher order hydrodynamic quantities. In such a way one could precisely relate the modifications~\eqref{tau3} in the shear mode at order $q^4$ to explicit hydrodynamic coefficients.

One possible extension of this work is to study how the CS coupling affects the gravitational and scalar perturbations of charged AdS black holes and the linear response of the holographic theory. One possible application in this context is, for example, including parity violation terms in holographic superconductors~\cite{Hartnoll:2008vx}. 
Also in this case scalar and gravitational perturbations will be coupled through the CS coupling and this may affect the superconducting phase and the electrical conductivity which, in the context of AdS/CFT correspondence, is computed from the Maxwell perturbations, coupled to gravitational perturbations. Again, the generalized series expansion can be used in this case.

Another extension would be considering higher dimensions. For instance, in five dimensions, the CS term for gravity reads
\be
\int \phi \epsilon_{abcde}R^{ab} \wedge R^{cd}\wedge e^ e d^ 5x,
\ee
where $R^{ab}$ is the curvature form and $e^a$ is the f\"unf-bein. In principle, this term should not be topological anymore, in the same way that the Gauss-Bonnet term is not a topological term in higher dimensions. It is naturally expected that spherically symmetric solutions to GR are still embeddable in DCS gravity. In this case, the dual theory is better known and the scalar operator in the dual theory can be explicitly identified.
\section{Acknowledgement}
This work was supported by the {\it DyBHo--256667} ERC Starting and by FCT - Portugal through projects PTDC/FIS/098025/2008,
PTDC/FIS/098032/2008, PTDC/CTE-AST/098034/2008 and CERN/FP/116341/2010. Computations were performed on the TeraGrid clusters TACC Ranger and NICS Kraken, the Milipeia cluster in Coimbra, Magerit in Madrid and LRZ in Munich.
\appendix
\section{Spherical Black Hole}
%
\begin{figure*}[htb]
\begin{center}
\begin{tabular}{cc}
\epsfig{file=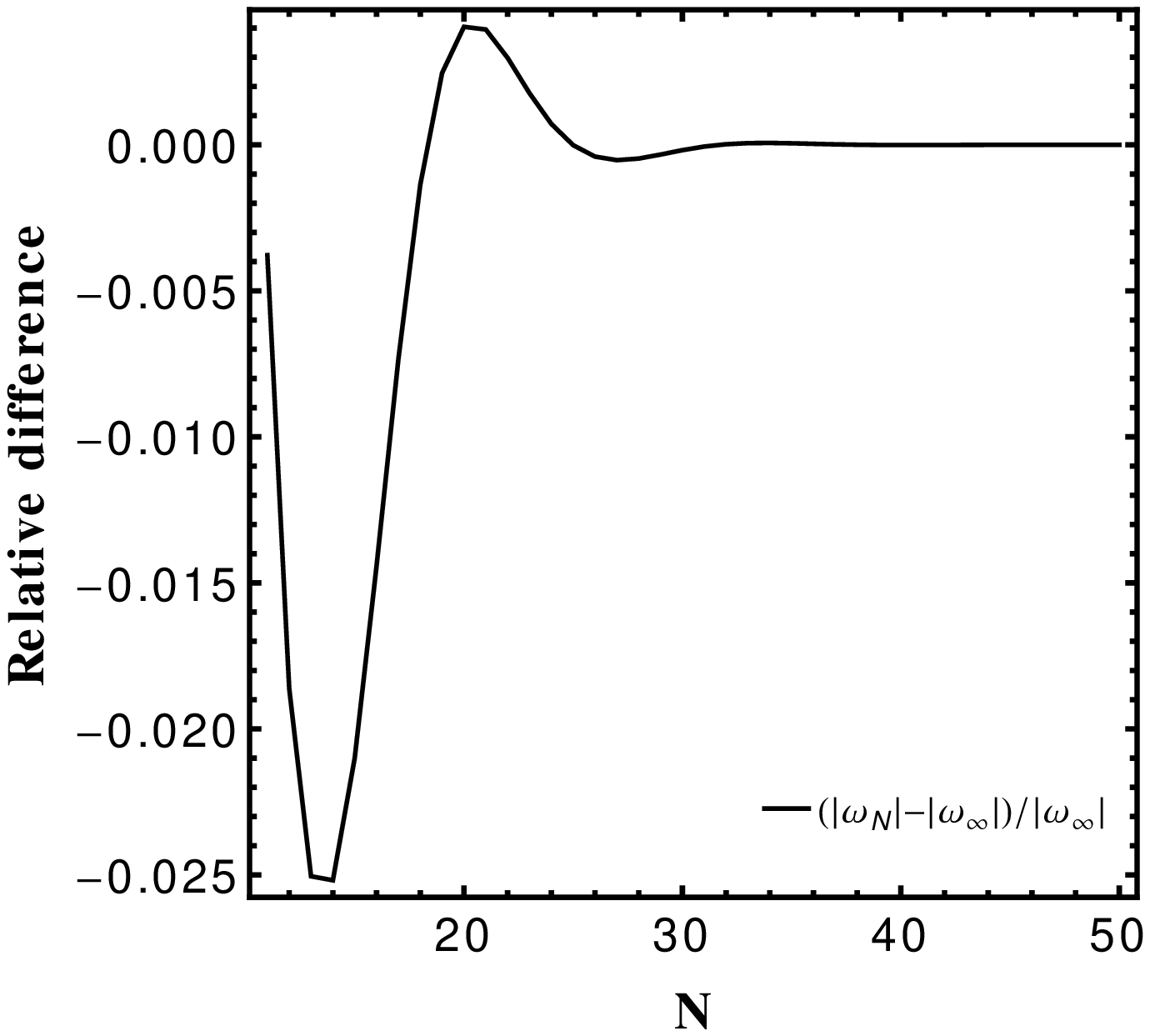,height=6.5cm,angle=0}&
\epsfig{file=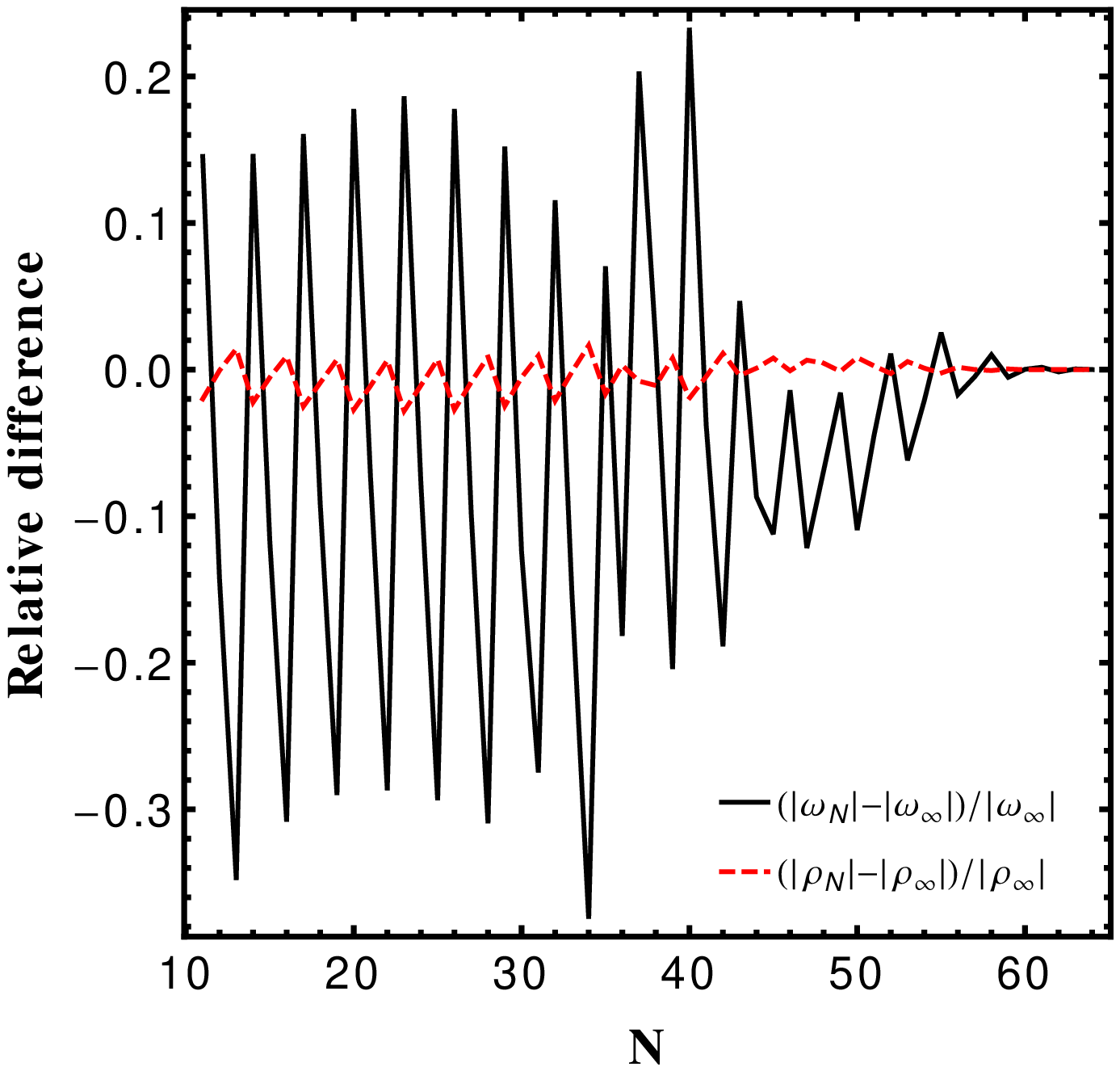,height=6.5cm,angle=0}
\end{tabular}
\caption{Convergence plot for $\alpha=0$ (left panel) and $\alpha/r_h^2=1$ (right panel).
In the left panel we show the relative difference between the QN frequencies computed with a truncation order $N\to\infty$ and those computed at a given $N$, $(|\omega_N|-|\omega_\infty|)/|\omega_\infty|$. In the right panel we also show the same convergence plot for the ratio $\rho=\sigma_0/q_0$.
\label{fig:conv_sph}}
\end{center}
\end{figure*}
\begin{figure*}[htb]
\begin{center}
\begin{tabular}{ccc}
\epsfig{file=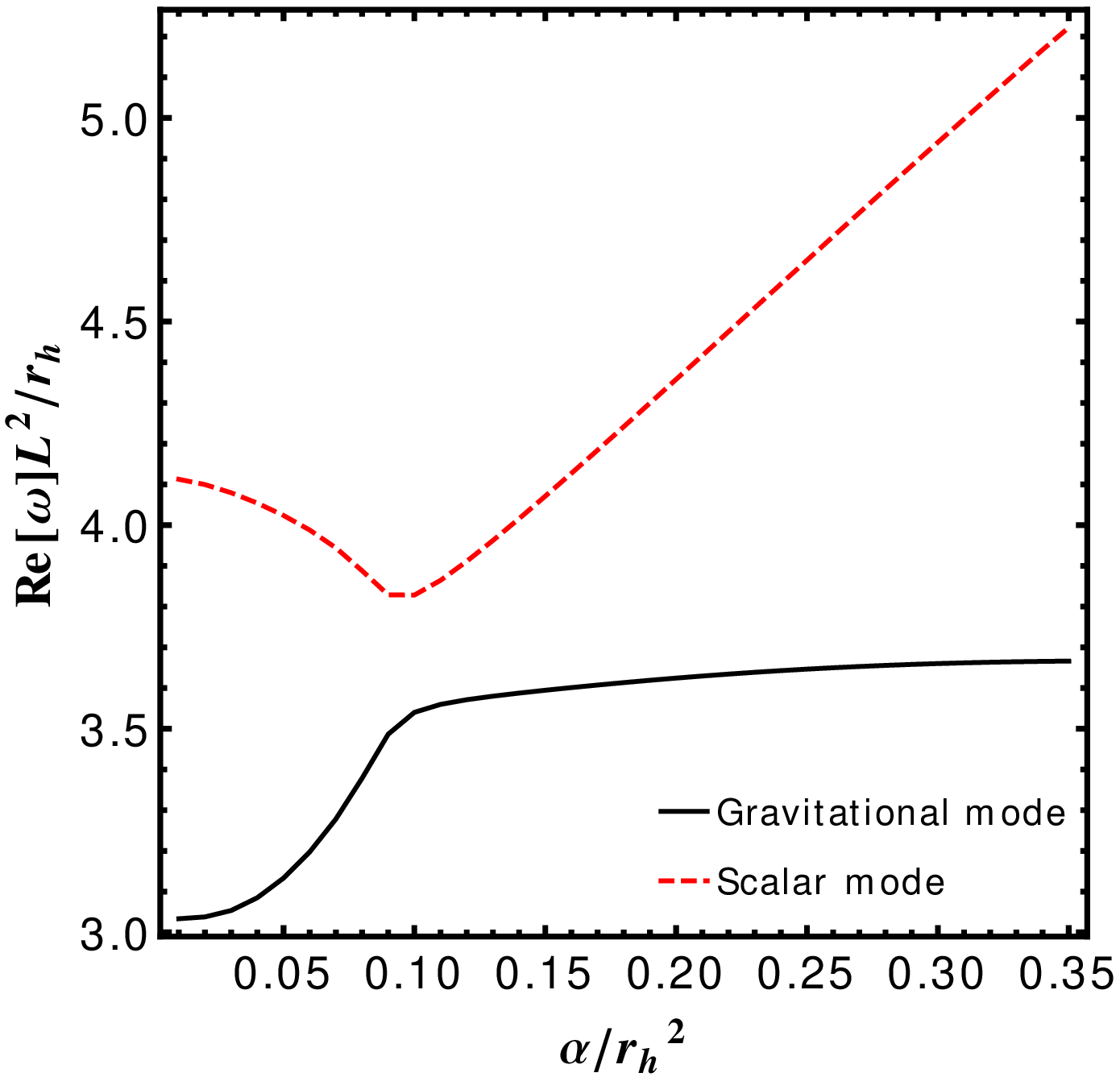,height=5.2cm,angle=0}&
\epsfig{file=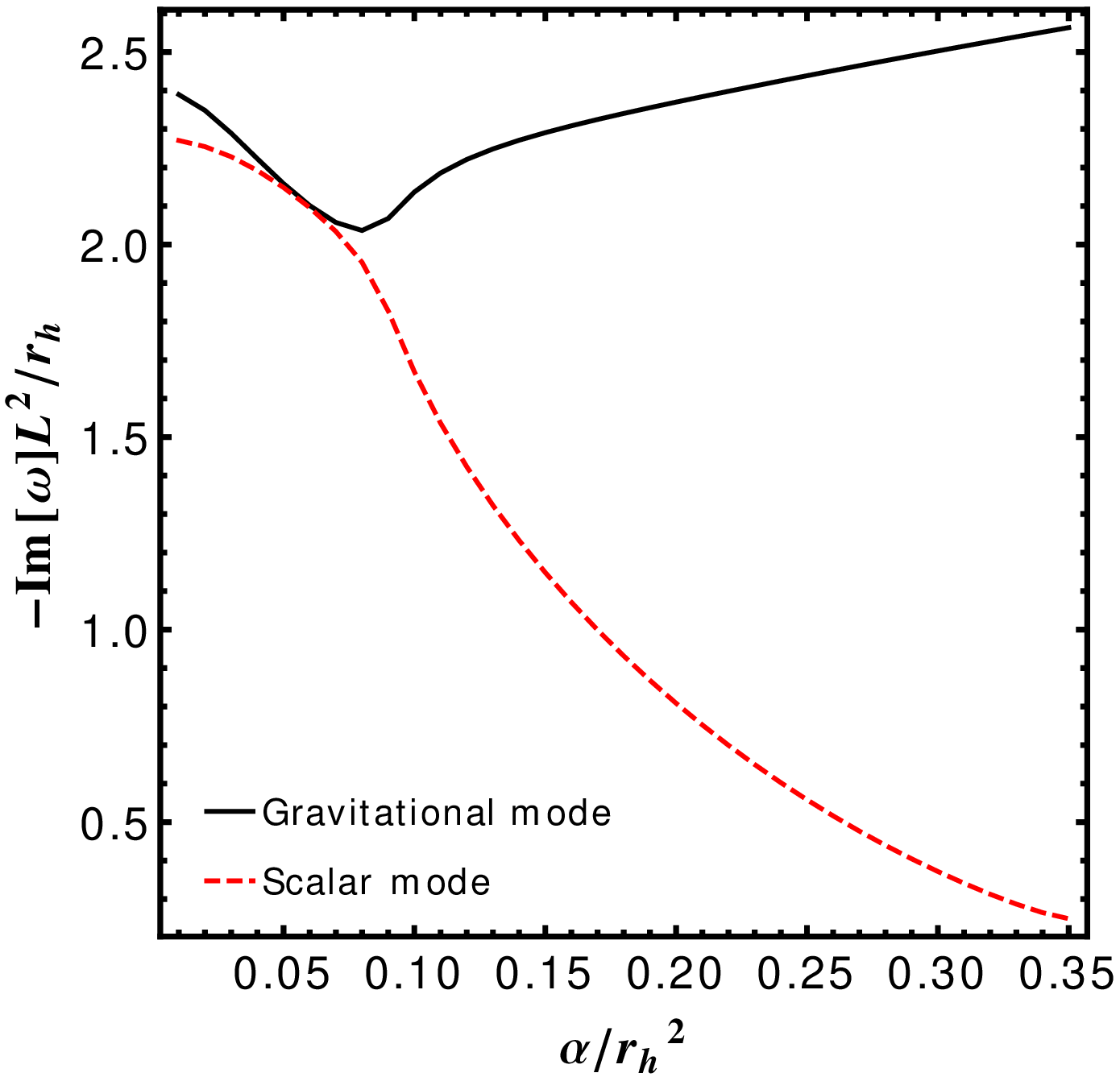,height=5.2cm,angle=0}&
\epsfig{file=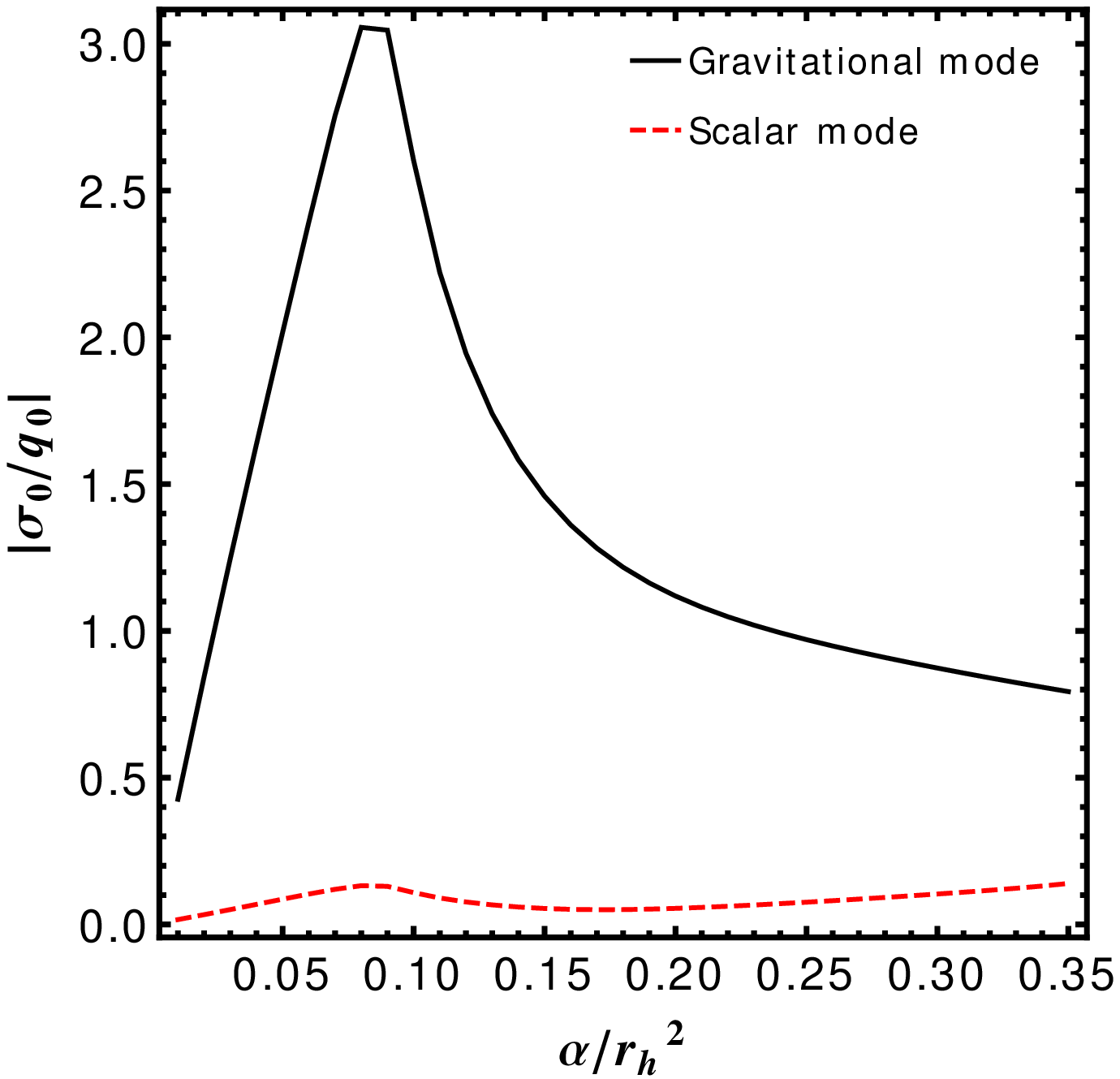,height=5.2cm,angle=0}
\end{tabular}
\caption{The quasinormal frequencies as function of DCS coupling $\alpha=0$ and (right panel) the norm of the ratio $\sigma_0/q_0$.
\label{fig:freq}}
\end{center}
\end{figure*}

For completeness, in this appendix we report our results for the spherically symmetric black hole ($\kappa=1$). We investigated how the first overtone mode is changed by the CS coupling constant for small black holes, namely $r_h/L=1$ and $q=2$.
Fig.~\ref{fig:conv_sph} shows how the results converge as function of the truncation order $N$ as $\alpha$ increases, demonstrating a need for larger truncation numbers as the coupling increases. As we discussed for the planar case, our results exhibit a mixing between the the gravitationally sector to the scalar sector.  In fact, both modes exist for $\alpha=0$ where the equations are decoupled; in this sense, the modes are either purely gravitational, either purely scalar. Once $\alpha$ is set on, the modes start to mix, but in regimes where $\sigma_0/q_0\ll1$ ($\sigma_0/q_0\gg1$), the mode is dominated by the gravitational (scalar) sector. In this sense, the mode originating from the scalar sector in the $\alpha=0$ theory stays dominated by the scalar sector,whereas the mode originating from the gravitational sector is rapidly mixed with the scalar mode in the sense that the ratio $\sigma_0/q_0$ becomes of order $1$. These considerations can be appreciated in Fig.~\ref{fig:freq}, where the real and imaginary parts of the modes are shown, together with the ratio $\sigma_0/q_0$.
%
%
%
%

%

\end{document}